\documentclass[12pt]{iopart}

\usepackage[sort&compress,numbers]{natbib}
\usepackage{xr-hyper}
\usepackage{graphicx}
\usepackage[dvipsnames]{xcolor}
\usepackage{subfig}
\usepackage[font=small,labelfont=bf]{caption}
\usepackage{hyperref}
\graphicspath{ {./images/} }

\begin{document}
\title{Machine-learning accelerated identification of exfoliable two-dimensional materials}

\author{Mohammad Tohidi Vahdat$^1$$^,$$^2$, Kumar Agrawal Varoon$^2$, and Giovanni Pizzi$^{1,3,*}$}

\address{$^1$ Theory and Simulation of Materials (THEOS) and National Centre for Computational Design and Discovery of Novel Materials (MARVEL), EPFL, Lausanne, Switzerland}
\address{$^2$ Laboratory of Advanced Separations (LAS), École Polytechnique Fédérale de Lausanne (EPFL), Switzerland}
\address{$^3$ Laboratory for Materials Simulations (LMS), Paul Scherrer Institut (PSI), CH-5232 Villigen PSI, Switzerland}
\address{$^*$ Author to whom any correspondence should be addressed.}
\ead{giovanni.pizzi@epfl.ch}

\begin{abstract}
Two-dimensional (2D) materials have been a central focus of recent research because they host a variety of properties, making them attractive both for fundamental science and for applications. It is thus crucial to be able to identify accurately and efficiently if bulk three-dimensional (3D) materials are formed by layers held together by a weak binding energy that, thus, can be potentially exfoliated into 2D materials.
In this work, we develop a machine-learning (ML) approach that, combined with a fast preliminary geometrical screening, is able to efficiently identify potentially exfoliable materials.
Starting from a combination of descriptors for crystal structures, we work out a subset of them that are crucial for accurate predictions. Our final ML model, based on a random forest classifier, has a very high recall of 98\%. Using a SHapely Additive exPlanations (SHAP) analysis, we also provide an intuitive explanation of the five most important variables of the model. Finally, we compare the performance of our best ML model with a deep neural network architecture using the same descriptors.
To make our algorithms and models easily accessible, we publish an online tool on the Materials Cloud portal that only requires a bulk 3D crystal structure as input. Our tool thus provides a practical yet straightforward approach to assess whether any 3D compound can be exfoliated into 2D layers.
\end{abstract}

\noindent{\it Keywords:\/} Two-dimensional materials, Exfoliation, Crystal structure, Binding energy, Online tool

\maketitle

\section{Introduction}
2D materials have been extensively explored for various applications across several disciplines~\cite{Xia2014,Fiori2014,Saito2016,Manzeli2017,Mannix2017, Vahdat2020,Vahdat2021}. In part, the scientific excitement about 2D materials originates from their ultimate thinness, making them exceptionally promising for applications, e.g., in electronics~\cite{Radisavljevic2011,Chhowalla2016, Pizzi2016}, in catalysis~\cite{Wang2016,Deng2016,Zhu2018}, or in molecular separations ~\cite{Kumar2012,Li2013,Peng2014}. Moreover, the physical properties of monolayers are often distinct from those of their parent 3D materials, since charge and heat transports are restricted to a plane~\cite{Butler2013}. Therefore, this makes them attractive not only for applications, but also for fundamental science \cite{Moore2010,Kou2014,Walsh2017,Marrazzo2018}. 
Furthermore, the possibility of realizing stacks of 2D layers into van-der-Waals (vdW) heterostructures opens the way to a combinatorially large number of novel properties to be explored~\cite{Nayak2017,Tran2019,Alexeev2019,Pizzi2021,Marrazzo2021}.

Decades ago, it was demonstrated that layered van-der-Waals materials, such as metal dichalcogenides, can be mechanically and chemically exfoliated into a few or even single layers~\cite{Frindt1966}.
Since the experimental demonstration of the exfoliation of graphene~\cite{K.S.Novoselovetal2004}, the interest in these materials and in the experimental approaches to exfoliate them has sparked \cite{Kaniyoor2010,Coleman2013,Yi2015,Magda2015}.
However, experimentally synthesizing these materials is time-consuming, only a fraction of them have ever been synthesized, resulting in a limited diversity of the 2D materials that are investigated experimentally. 
First-principle simulations, and in particular density-functional theory (DFT) calculations, have been thus recently employed to determine the binding-energy strength between layers and consequently assess whether a structure can be exfoliated, resulting in the creation of databases of potential 2D materials \cite{Lebegue2013, Ashton2017,Cheon2017,Choudhary2017,Mounet2018,Haastrup2018,Zhou2019}. 
However, such calculations are quite expensive and require deep understanding of the simulation methods to obtain accurate results, as well as the capability of using a workflow infrastructure to manage the resulting large number of calculations \cite{Adhianto2010,Ong2013,ase-paper,Huber2020}.
Therefore, first-principles simulations might be a barrier when needing a fast, yet accurate, prediction of the exfoliability of candidate materials.

\begin{figure*}[t]
\centering
\includegraphics[width=15cm]{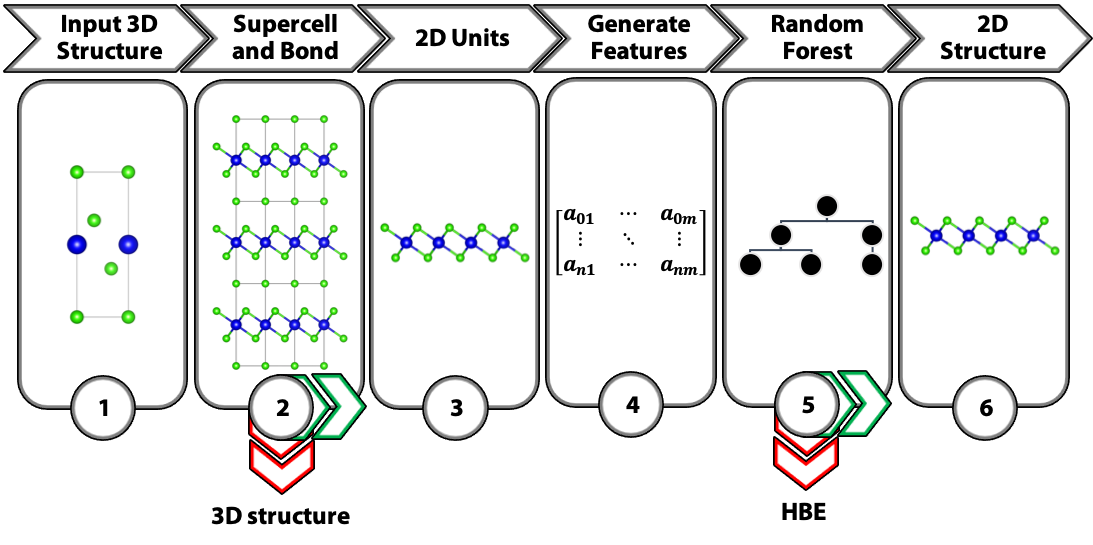}
\caption{Schematic illustrating the steps of our algorithm to efficiently determine the exfoliability of 3D materials. From left to right: 1. The procedure initiates from the unit cell of a bulk crystal structure. 2. A $3\times 3\times 3$ supercell of the bulk primitive cell is constructed, and chemical bonds are identified by comparing the bond distances with the atomic vdW radii. 3. Low-dimensional subunits are determined based the rank of the lattice vectors embedded into chemically bonded units, and a 2D primitive cell is built (\emph{LowDimFinder} geometrical pre-filtering step). 4. The crystal structures passing the previous steps are represented with numeric features, used as input of a random forest classifier (step 5) trained and tested on a set of over 2000 binding energies calculated with DFT. Our ML model predicts if the material has a high binding energy (HBE) and should be discarded, thus filtering out potential 2D structures for further analysis (step 6).}
\label{fig:workflow}
\end{figure*}

Here, we develop a combined geometrical and ML model that can provide accurate and fast predictions on whether a bulk 3D material is formed by layers held together by a weak binding energy and, thus, is potentially exfoliable into 2D layers.
Our multi-step procedure, illustrated in Fig.~\ref{fig:workflow}, starts by pre-screening layered structures based on geometrical criteria requiring only the atomic positions of the atoms in the structure.
The resulting filtered structures are featurized, and finally a ML model based on a random forest classifier is applied to assess whether the material can be exfoliated or, instead, has a high binding energy (HBE).

We train and then evaluate the performance of the model using a set of 2392 structures from Ref.~\cite{Mounet2018}, obtained from the preliminary geometrical step applied on a large initial set of 108,423 unique 3D structures originating from the ICSD~\cite{ICSD} and COD~\cite{Grazulis2012} databases.
To capture the underlying structural patterns that correlate with the exfoliability of a material, we first reduce the number of descriptors needed by the model, and then use a SHapely Additive exPlanations (SHAP) analysis to explain the most relevant features.
We also compare our best-performing ML model with a deep neural network.
Finally, in order to make our methodology and code immediately available to all researchers without the need of any software installation, we implement and publish an online web tool on the Materials Cloud web platform~\cite{talirz_materials_2020}, available at \url{http://ml-layer-finder.materialscloud.io} and working directly in the web browser.
The tool provides the opportunity, even for non-experts of simulations, to just upload a crystal structure in a variety of file formats and rapidly evaluate if it can be exfoliated into 2D layers or not.

\section{Methods}
\subsection{Description of the crystal structures and the method used to  identify layered compounds}
In the following, we will indicate exfoliable 2D materials as a positive hit of a model, while a HBE material as a negative hit.
Therefore, false positives indicate HBE systems that are instead predicted to be exfoliable (2D), while false negatives indicate exfoliable systems that the model suggests to discard as HBE.

Our screening procedure (steps 1 and 2 of Fig.~\ref{fig:workflow}) starts from the comprehensive initial set of 108,423 unique bulk 3D crystals extracted from ICSD and COD, used in the study by Mounet \emph{et al.}~\cite{Mounet2018,Mounet2020}.
In Ref.~\cite{Mounet2018}, the preliminary geometric step (that we will refer to as the \emph{LowDimFinder}) identifies chemically-bonded subunits by comparing interatomic distances $d_{AB}$ between pairs of atoms $A$ and $B$ in the structure with the corresponding atomic van-der-Waals radii $r_A$ and $r_B$ (radii are taken from from Ref.~\cite{Alvarez2013}).
We consider two atoms as bonded if $d_{AB} \leq r_A + r_B - \Delta$, where $\Delta$ is a tunable parameter (see below).
The dimensionality of each connected manifold is then obtained from the rank of the set of 3D lattice vectors embedded into the manifold (see ref.~\cite{Mounet2018} for more details).
One main advantage of this approach is identifying subunits of any dimensionality (i.e., 2D layers, 1D chains, and 0D clusters). Moreover, in the case of 2D layers, no specific orientation is assumed for the 2D plane with respect to the Cartesian axes.

For the purposes of preliminary filtering, we wish to have a very low rate of false negatives (i.e., a very high recall rate) so as not to miss important materials, while at the same time we can accept a moderate amount of false positives, because these can be further filtered later (first by the ML model discussed here, and possibly later by accurate DFT simulations following the application of our ML model).
Therefore, also considering that this first filtering is only aware of distances between atoms and not of their chemistry (except for the atomic vdW radii), we consider five values for the tunable parameter, namely $\Delta= 1.1, 1.2, 1.3, 1.4, 1.5$~\AA.
Any crystal structure recognized as 2D for at least one of the five values of $\Delta$ is then handed to the ML model in the following steps.

\subsection{Choice of descriptors for 3D compounds}
A crucial preliminary step before implementing and training a ML model is to represent crystal structures in terms of numerical features representing the material chemistry and geometry.

Different strategies have been developed in the literature~\cite{Ward2016, Meredig2014, Bartok2013, De2016}.
In this work, all the descriptors are chosen from those available and implemented in the matminer package~\cite{Ward2018}. Choosing the right descriptors requires care and testing, and is dependent on the details of the descriptors and the properties of interest. As such, there is not a single unique correct way to represent crystal structures.
We start from a set of potentially relevant features and, as a first step, we identify which of these are most relevant (see Supplementary Information Sec. 1 for details).
Our final set of features are based on those proposed by Ward \emph{et al.}~\cite{Ward2017}, where space is partitioned into Wigner–Seitz cells using Voronoi tessellation.
These cells (including all points in space closer to a central atom than to any other atom) are then used to assign neighbors and then construct local descriptors of the environment that are only weakly sensitive to small changes that might occur during a geometry relaxation.
Derived from the Voronoi tessellation, several attributes are proposed in Ref.~\cite{Ward2017}. 
In the following we will restrict to three different sets of features that we selected for our final model: chemical ordering, maximum packing efficiency, and local environment attributes.
The first one, chemical ordering, describes how much the placement of species in a structure deviates from random, using Warren-Cowley-like ordering parameters \cite{Cowley1950}. The second one, maximum packing efficiency, measures the largest sphere that fits inside each Voronoi cell. The last, local environment attributes, are based on comparing elemental properties (such as the electronegativity and 21 other elemental properties) of the central atom to its neighbors, weighted by the surface area of the faces of the Wigner–Seitz cells. More technical details can be found in Ref.~\cite{Ward2017}. 

\subsection{Model selection and training}

To achieve a ML model that can classify potentially exfoliable materials, we train the model on binding energies for the 2392 structures prefiltered by the \emph{LowDimFinder}, that were computed using DFT in Ref.~\cite{Mounet2018}.
We first center the features at zero and rescale them using their mean and standard deviation, respectively. 
In order to determine the quality of the model on an unseen data set, train-test splitting is performed randomly, and the size of the train set is 70\%. All the statistics reported later are computed by averaging over ten different random seeds used for train-test splitting.

We tested various learning approaches, including deep learning architectures. Although some of the architectures perform well (see Supplementary Information Sec. 2), we decided to employ and discuss further in the following a random forest classifier ML model~\cite{Breiman2001}, as it is relatively simple and, at the same time, it demonstrates an excellent performance for our goal.

\section{Discussion and Results}

\begin{figure}[t]
\centering
\includegraphics[width=\linewidth]{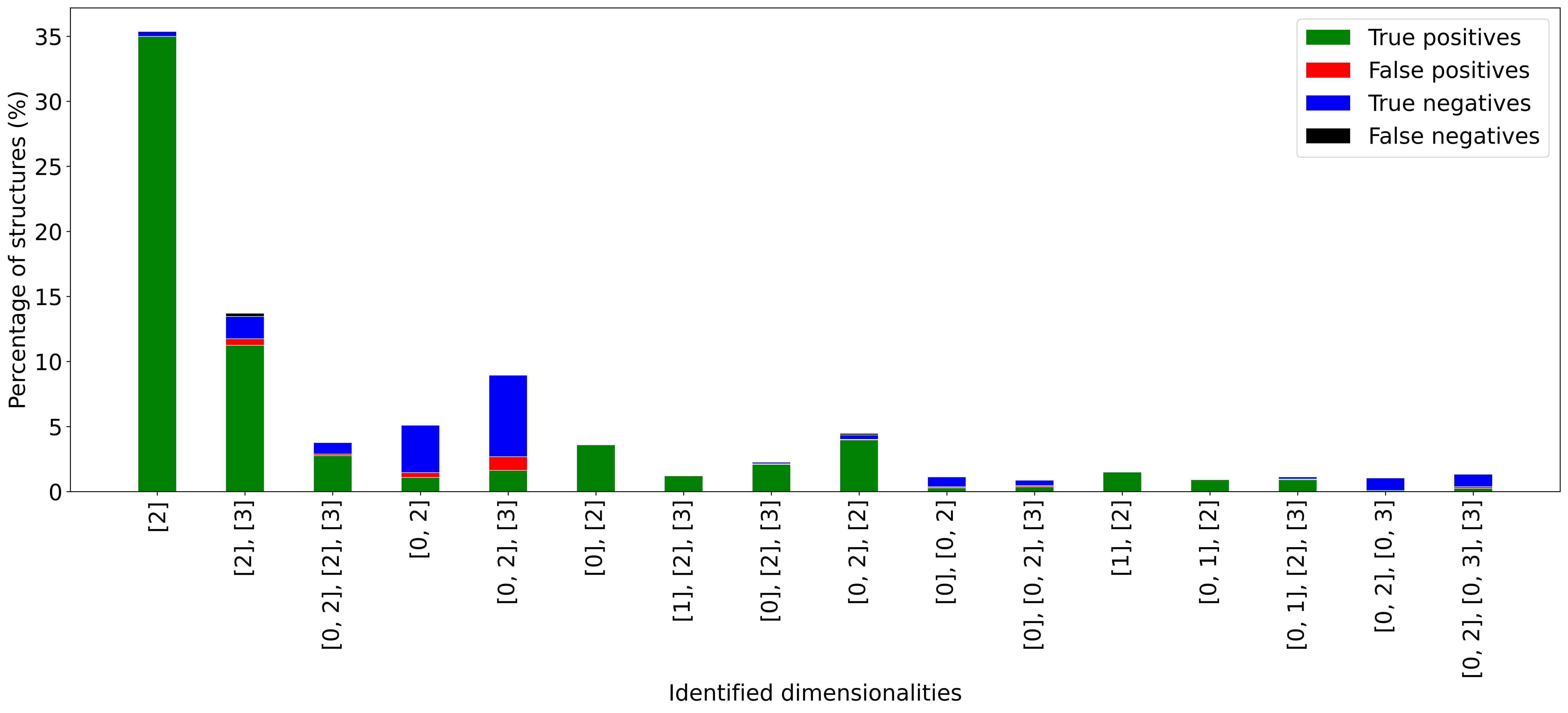}
\caption{Evaluation of the performance of our ML-based model with respect to accurate results on the binding energies calculated using DFT. 
The bar labels combine the dimensionalities obtained for all values of $\Delta$ used to determine atom bonds in the preliminary geometrical \emph{LowDimFinder} screening (see main text). For example, if the obtained dimensions of subunits for the five values of $\Delta$ are [0,2], [0,2], [0,2], [2] and [2], the resulting label reported here is [0,2],[2].
Structures are color-coded based on the results of the ML classifier. Green: True positives (i.e., structures that are correctly recognized as exfoliable by the ML model). Red: False positives (i.e., compounds incorrectly identified as exfoliable 2D). Blue: True negatives (i.e., structures correctly determined as HBE structures). Black: False negatives (i.e., structures incorrectly classified as HBE structures). The value on the $y$ axis is the percentage of structures with respect to the total number of structures with at least one 2D subcomponent in the \emph{LowDimFinder} analysis.}
\label{fig:Barplot-detail_ML}
\end{figure}

\begin{figure}[t]
\centering
\includegraphics[width=15cm]{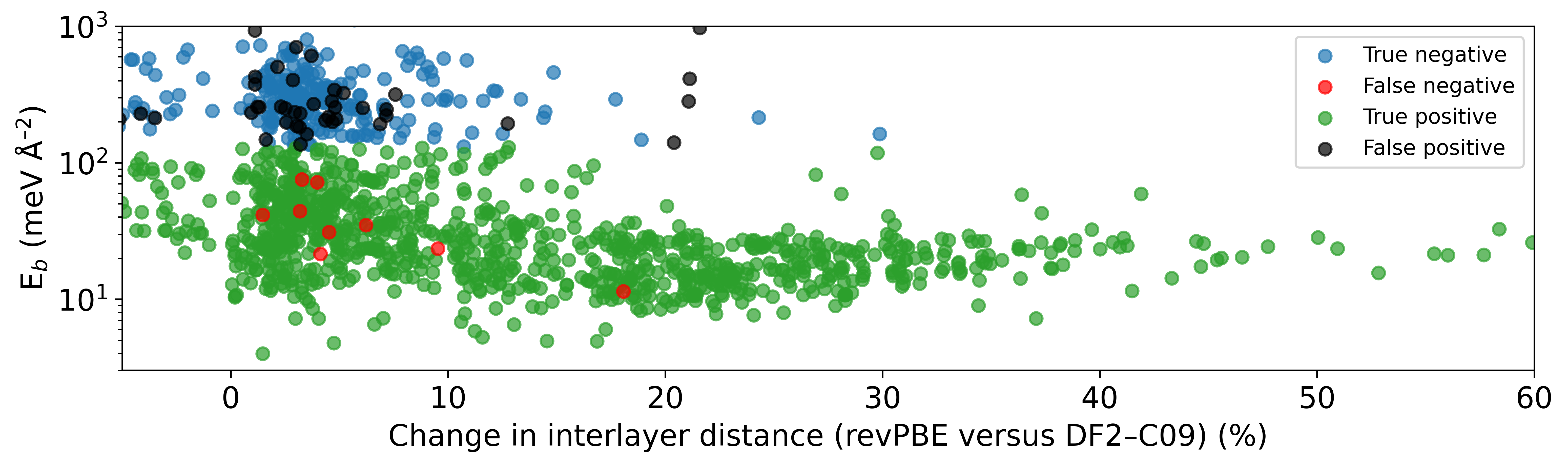}
\caption{Binding energy versus change in the interlayer distance (for the 3D structures relaxed using the vdW-aware DF2-C09 functional with respect to the relaxation with the revPBE functional, as defined in Fig. 2c of Ref.~\cite{Mounet2018}). Using our trained ML model, materials are classified as exfoliable or with high binding energy and reported in different colors. Green and blue circles show structures that are classified correctly both with vdW-DFT and our ML model as exfoliable and high binding energy structures (true positives and true negatives, respectively). Red circles indicate exfoliable structures according to vdW-DFT, but tagged as HBE by our ML model (false negatives), while black circles represent HBE structures as determined by vdW-DFT, while the ML model tags them as exfoliable (false positives).}
\label{fig:fig2C}
\end{figure}

Mounet \emph{et al.}~\cite{Mounet2018} performed binding energy calculations on 2662 prospective layered structures and identified those that are held together by weak interactions and are ready for exfoliation. 
We consider 2392 structures out of 2662 (two different vdW DFT functionals are used in Ref.~\cite{Mounet2018}; here, we restrict only to those for which binding energies are calculated using the vdW-DF2-C09 functional~\cite{Lee2010,Cooper2010,Hamada2010}, so as to have a consistent calculation method of binding energies for all structures in the dataset).
In particular, our portfolio contains 1813 structures that are determined to be exfoliable into 2D layers by DFT simulations including vdW interactions (vdW-DFT), defined as those whose binding energy is lower than 130 meV{\AA}$^{-2}$, while 579 structures have HBE (although the first \emph{LowDimFinder} preliminary filtering had not discarded them).

Fig.~\ref{fig:Barplot-detail_ML} illustrates the combined performance of the preliminary \emph{LowDimFinder} geometrical filtering followed by our ML model, with respect to the accurate vdW-DFT calculations, resolved with respect to the set of dimensionalities of the sub-units identified by \emph{LowDimFinder} (the performance of \emph{LowDimFinder} alone can be better inspected in Supplementary Fig.~\ref{fig:SI-Barplot-detail}).
In particular, in Fig.~\ref{fig:Barplot-detail_ML} we see that systems that are tagged by \emph{LowDimFinder} as only having 2D subcomponents for any value of $\Delta$ (labeled as \texttt{[2]}) are those with highest probability of displaying low binding energy, and thus promising for exfoliation.
Moreover, this set also has a very low rate of false positives. 
On the other hand, structures marked also displaying a zero-dimensional subunit for at least one value of $\Delta$ (and in particular \texttt{[0,2]}  and \texttt{[0,2],[3]}) include a relatively large fraction of systems that are confirmed to have HBE by vdW-DFT. 
Nevertheless, these results demonstrate that, while the dimensionalities found by \emph{LowDimFinder} alone can provide an indication of the probability of finding false positives, they are not sufficient (alone) to provide a reliable method to reduce their number without also increasing significantly the rate of false negatives (we remind that our design strategy is to have the smallest amount possible of false negatives, but avoiding to miss important 2D candidates).

\begin{table}[t]
\caption{\label{table:result}Performance of our ML model (based on a random forest classifier) on the test data set. Our model captures 513 out of 525 structures correctly (true positives) resulting in a very high recall rate of 98\% (i.e., the model has very few false negatives). Additionally, it reduces the number of false positives by $\approx 50$\% with respect to the initial \emph{LowDimFinder} geometrical screening.}
\footnotesize\centering
\begin{tabular}{@{}lrrrr}
    &&&&\textbf{Total number}\\
    \textbf{Label}&\textbf{Precision}& \textbf{Recall}&\textbf{F1-score}&\textbf{of structures}\\
\br
\textbf{HBE}&85\%&49\%&62\%&139\\
\mr
\textbf{2D}&88\%&98\%&93\%&525\\
\end{tabular}\\

\end{table}

Therefore, we train our ML model (based on the random forest classifier) to improve the classification of structures in 2D and HBE, still at a negligible fraction of the cost of a full DFT simulation. 
As visible in Fig.~\ref{fig:Barplot-detail_ML}, the ML model is able to correctly label most HBE systems that had passed the \emph{LowDimFinder} prefilter (true negative systems, in blue). Detailed results on the performance of our ML model is reported in Table~\ref{table:result}. In the test dataset, we have 664 structures, including 139 HBE structures and 525 exfoliable structures.
Our model managed to correctly predict 513 structures as true positives (i.e., a recall of 98\%, very close to our ideal goal of 100\%), while at the same time reducing the number of false positives of the \emph{LowDimFinder} by about 50\%.
Therefore, our model reaches our goal of significantly filtering the number of candidates that can go to an ultimate screening step using DFT, thus reducing the associated computational cost.

\begin{figure*}[t]
\centering
\includegraphics[width=12cm]{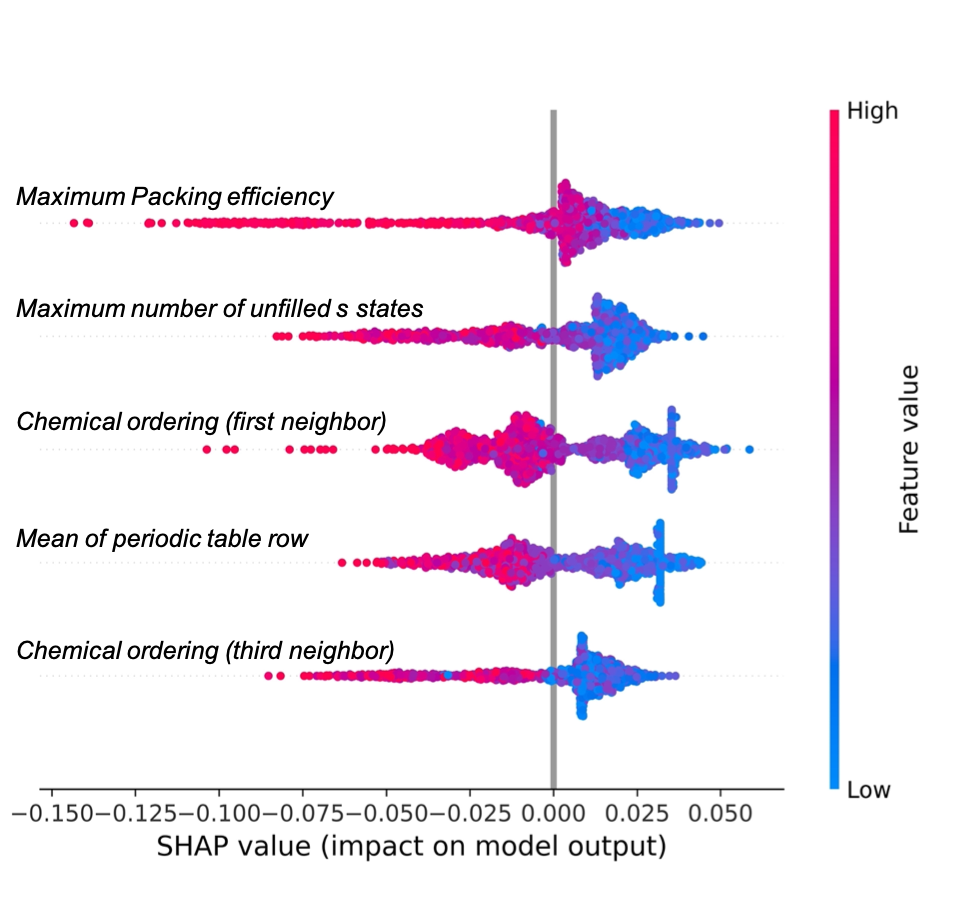}
\caption{SHAP plot summarizing the five most important variables for our model, evaluated on all structures in the dataset. The color corresponds to the value of each input variable and reveals if there is a positive or negative correlation with the possibility of exfoliation (e.g., red circles on the left-hand side of the plot indicate a negative correlation). Additionally, the violin plots show the distribution of the importance, i.e., the spread of the SHAP values (along the $x$ axis) and how many samples have similar SHAP values (the thickness of the violin).}
\label{fig:features}
\end{figure*}

\begin{figure}[t]
\centering
\includegraphics[width=12cm]{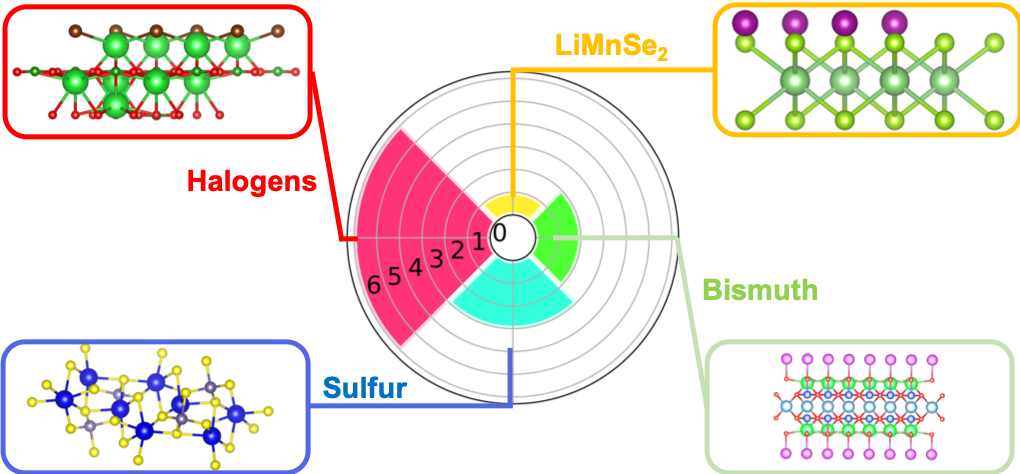}
\caption{Summary of the 12 false negatives of our ML model (out of the 525 exfoliable structures in our test set). In addition to LiMnSe$_2$, 6 structures include halogens, 3 include sulfur and  2 include bismuth.}
\label{fig:false-negatives}
\end{figure}

To better understand the distribution of false positives and false negatives of our ML model, we first show in Fig.~\ref{fig:fig2C} the binding energies for all compounds included in our training set, as a function of the change in the interlayer distance with two DFT functionals (including or not vdW interactions, respectively, as an approximate proxy for the strength of vdW interactions in the material).
This figure mirrors Fig. 2c of Ref.~\cite{Mounet2018}, while adding information on the performance of our ML model.
The figure visualizes graphically the model performance, and shows that false positives and negatives do not have a specific distribution with respect to the two variables of the scatter plot.

We then use the SHAP technique~\cite{NIPS2017_7062} to gain a deeper understanding of what the model learned in the training process. 
SHAP can disclose how the features used by the model affect the predictions.
Fig.~\ref{fig:features} shows the five most important variables of our model: maximum packing efficiency, maximum number of unfilled s states, chemical ordering (first neighbor), mean of periodic table row, and chemical ordering (third neighbor). The importance here is defined as the mean absolute SHAP value of all the points in the dataset. The colors in the figure denote the value of the input variable, where red indicates high and blue indicates low values. Consequently, red on the right-hand (left-hand) side of the plot implies a positive (negative) correlation with exfoliation.

Fig.~\ref{fig:features} shows that maximum packing efficiency is the most important feature and has a negative correlation with exfoliation, consistent with what one can expect as this feature indicates the largest sphere that fits inside each Voronoi cell, and structures with high packing efficiency will be most probably not layered (packing ratio was already used as a filtering criterion in one of the early DFT studies to predict 2D materials~\cite{Lebegue2013}). 
The second most important feature is the maximum number of unfilled s states. Indeed, as the number of unfilled s states increases, the material will tend to form stronger bonds and will be consequently less prone to break them and be exfoliated, justifying its negative correlation.
Another important variable is chemical ordering (both for first and third neighbors). This variable is associated with the arrangement of the crystal structure, where structures with ordered arrangements will have a larger value (closer to 1), while random arrangements will have a value closer to zero.
Also in this case, the negative correlation visible in Fig.~\ref{fig:features} can be expected, because structures very ordered configurations are typically strongly bonded 3D materials that are not composed by exfoliable layers. 
Finally, another important feature is the mean of the periodic table row, indicating a tendency of lighter elements (i.e., small value for the periodic table row) to form exfoliable materials.

Finally, since our aim is to reduce as much as possible the number of false negatives, we report explicitly the 12 exfoliable structures that are considered as HBE by our ML model. These are schematically represented in Fig.~\ref{fig:false-negatives} (the complete list is shown in Supplementary Information Table \ref{table:False-negatives}). 
We highlight in particular that 6 structures out of 12 contain halogen elements, and 3 contain sulfur.
This result can be used as a guide for further improvements of the model, e.g. to include further descriptors that can help in correctly identifying structures with low binding energy (we note, however, that other structures including the same chemical elements exist, where the model works correctly: therefore their sole presence cannot be used as a simple criterion to identify false negatives).

\section{Online tool}
In order to give easy access to our model and its predictions without the need of preliminary expertise, we also implement an online web tool that we publish on the Materials Cloud web platform \cite{talirz_materials_2020} at the URL \url{http://ml-layer-finder.materialscloud.io}. The tool works directly in the browser and does not require any installation.

\begin{figure}[t]
\centering
\includegraphics[width=12cm]{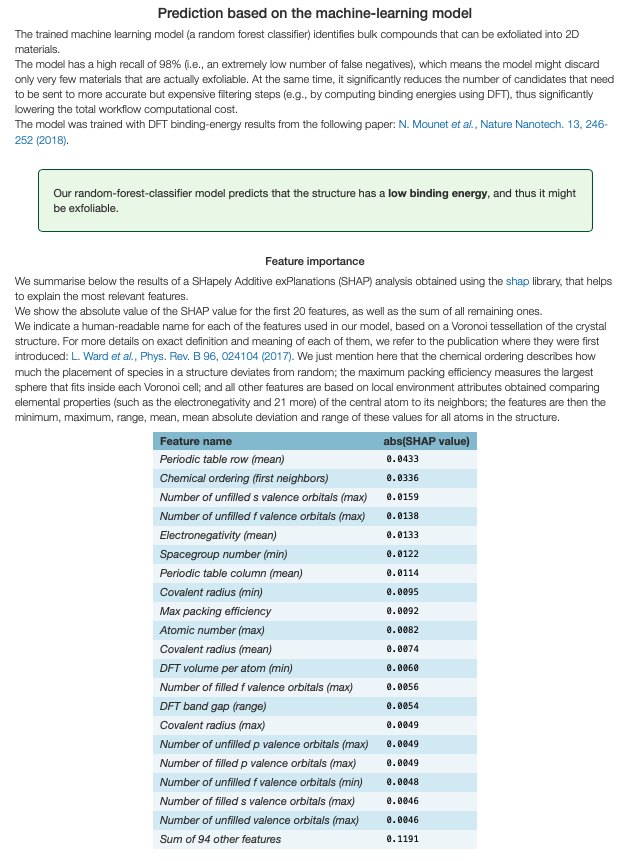}
\caption{Screenshot of the online tool implementing the algorithms of this paper, available on the Materials Cloud \cite{talirz_materials_2020} Work/Tools section. Once a material is uploaded, the output page displays the prediction of exfoliability by our ML model. Here, a screenshot of a part of the tool results page is shown, displaying the ML model prediction in the case of graphite. The output page of the tool includes more information, such as the visualization of the crystal structure and its decomposition into 2D layer components.}
\label{fig:OnlineTool}
\end{figure}

In the first selection page, the user can upload a bulk crystal structure. Several common formats are supported thanks to the parsers implemented in the ASE~\cite{hjorth_larsen_atomic_2017} and pymatgen~\cite{ong_python_2013} libraries. Once the bulk structure is uploaded (or selected from already available examples), the tool performs the needed computations to first determine if the material is layered, and then to predict if it is exfoliable.
In particular, the tool (following the workflow of Fig.~\ref{fig:workflow}) first computes the bonds and then identifies the dimensionality of each connected sub-manifold using the \emph{LowDimFinder}.
Any subunit that is a 2D layer is displayed on the output web page.
The tool then generates feature vectors for these structures, and then it uses the pre-trained ML model to predict if the crystal structure can be exfoliated or has a HBE.
The result is provided on output, together with a table summarizing the the SHAP value of the most important features for the uploaded structure (see Fig.~\ref{fig:OnlineTool}).

\section{Conclusion}
We presented an efficient yet accurate method combining a preliminary geometrical screening followed by a random-forest-classifier ML model to identify bulk compounds that can be exfoliated into 2D materials. 
The procedure that we developed provides an effective and straightforward prefilter to assess whether any bulk structure has the potential to be exfoliated: thanks to its high recall of 98\%, our method has an extremely low rate of false negatives, i.e., of exfoliable materials that would be discarded by the model and thus be missed in a search for interesting candidates.
Yet, it helps in significantly reducing the number of candidates that need to be sent to more accurate but expensive filtering steps (e.g., by computing binding energies using DFT), thus significantly lowering the total workflow computational cost.
By means of a SHAP analysis, we also provide an explanation of the most important features used by our model to determine exfoliability. These include the maximum packing efficiency, the maximum number of unfilled s sates, the chemical ordering and the mean of the periodic table row.
Finally, in order to enable even non-expert fellows to run our model and predict exfoliability of new materials, we implemented and deployed it as an online web tool published on the Materials Cloud web platform, only requiring to provide the parent crystal structure as input. 

\section*{Data availability}
Supplementary information is available for this paper.
Data that were used and are needed to reproduce this study are deposited on the Materials Cloud Archive~\cite{talirz_materials_2020} and are available at \url{https://doi.org/10.24435/materialscloud:m4-7f}.

\section*{Code availability}
The code for parsing, featurization of structures and data analysis is available free of charge on Github at \url{https://github.com/Mohammad-vahdat/Machine_learning_accelerated_identification_of_exfoliable_two_dimensional_materials}.

\section*{Supplemental Information}
Supplemental Information for this article is available online.

\section*{Acknowledgment}
This research was supported by the NCCR MARVEL, a National Centre of Competence in Research, funded by the Swiss National Science Foundation (grant number 205602).
The authors gratefully acknowledge Nicola Marzari and Seyed Mohamad Moosavi for fruitful discussions, and Nicolas Mounet and Philippe Schwaller for the initial implementation of the \emph{LowDimFinder} code used in Ref.~\cite{Mounet2018}.

\section*{Supplementary Information}

\subsection*{SI: Features Selection}
Besides features mentioned in the main text, we tested other features, including structural heterogeneity (measuring the variances in the bond lengths and atomic volumes in a structure) and effective coordination number, which is based on the mean, maximum, minimum, and mean absolute deviation in the coordination number of each atom~\cite{Ward2017}. Moreover, composition-based features are generated based on the fractions of each element using the Magpie package~\cite{Ward2017,Ward2016}. The composition-based attributes tested in this work are the following:
\begin{itemize}
  \item \emph{stoichiometric attribute}, which depends on the fractions of each element (and not on which specific elements appear in the structure);
\end{itemize}

\begin{itemize}
  \item \emph{elemental-property-based attributes}, showing statistics of the elemental properties of all atoms in the crystal;
\end{itemize}

\begin{itemize}
  \item \emph{electronic-structure attributes}, which depend on the fraction of electrons in the $s$, $p$, $d$, and $f$ shells of the constituent elements, normalized by the total number of electrons in the system~\cite{Meredig2014};
\end{itemize}

\begin{itemize}
  \item \emph{ionicity attributes}, derived from differences in electronegativity between constituent elements.
\end{itemize}

However, in our evaluation step, we obtain that these features are not necessary to obtain accurate predictions, as they show small correlations to the output predictions. We therefore removed them from the final trained model discussed in the main text, to put more emphasis on the most critical and predictive attributes. The features used in the final model are described in the main text.

\subsection*{SI: Deep-learning architecture and training}

To compare the performance of our random-forest classifier model described in the main text with a deep neural network, we train fully-connected networks with three hidden layers and 128 neurons on each hidden layer. A ReLU activation function is used in the hidden layers. Dropout layers are added between each hidden layer to avoid overfitting. The last layer has one neuron to make the binary classification (2D vs.\@ HBE) with a sigmoid activation function. The network is implemented in TensorFlow 2~\cite{tensorflow2015-whitepaper} and trained for 46 epochs. The loss function is binary cross-entropy, which is optimized with the Adam optimizer~\cite{kingma2014method} with a learning rate of ${10}^{-3}$. Table \ref{table:result} summarizes the performance of the architecture.

\begin{table}[tb]
\caption{\label{table:result}Performance of the trained deep learning model on the test data set. The deep learning model captures 513 out of 525 structures correctly, while it reduces the number of false positives by 35 percent (structures that by mistake were considered as exfoliable).}
\footnotesize\centering
\begin{tabular}{@{}lrrrr}
&&&&\textbf{Total number}\\
\textbf{Label}&\textbf{Precision}& \textbf{Recall}&\textbf{F1-score}&\textbf{of structures}\\
\br
\textbf{HBE}&80\%&35\%&48\%&139\\
\mr
\textbf{2D}&85\%&98\%&91\%&525\\
\br
\end{tabular}\\

\end{table}

\begin{table}[tb]
  \caption{\label{table:False-negatives} The list of the 12 exfoliable structures that are considered as HBE by our random-forest classifier ML model (false-negative structures).}
  \footnotesize\centering
  \begin{tabular}{ccc}
  \multicolumn{3}{c}{\textbf{Experimental 3D parent structures}}                                  \\ \br
  \multicolumn{1}{c}{Formula}       & \multicolumn{1}{c}{Source database} & Database ID \\
   & (ICSD: \cite{ICSD}; COD: \cite{Grazulis2012}) & \\ \br
  \multicolumn{1}{c}{Ge$_3$Bi$_2$Te$_6$}     & \multicolumn{1}{c}{ICSD}            & 30394       \\ \mr
  \multicolumn{1}{c}{Tl$_{16}$S$_{12}$}       & \multicolumn{1}{c}{ICSD}            & 2647        \\ \mr
  \multicolumn{1}{c}{Ca$_4$Al$_2$Pb$_2$F$_{18}$}  & \multicolumn{1}{c}{COD}             & 9002898     \\ \mr
  \multicolumn{1}{c}{Sr$_2$CaCu$_2$Bi$_2$O$_8$} & \multicolumn{1}{c}{ICSD}            & 68188       \\ \mr
  \multicolumn{1}{c}{Na$_8$Ti$_4$S$_8$O$_4$}    & \multicolumn{1}{c}{ICSD}            & 67886       \\ \mr
  \multicolumn{1}{c}{LiMnSe$_2$}       & \multicolumn{1}{c}{ICSD}            & 50817       \\ \mr
  \multicolumn{1}{c}{Na$_6$Sc$_2$Br$_{12}$}    & \multicolumn{1}{c}{ICSD}            & 401335      \\ \mr
  \multicolumn{1}{c}{Ba$_4$B$_2$Br$_2$O$_6$}    & \multicolumn{1}{c}{ICSD}            & 190941      \\ \mr
  \multicolumn{1}{c}{Ba$_3$In$_2$Br$_2$O$_5$}   & \multicolumn{1}{c}{COD}             & 2002540     \\ \mr
  \multicolumn{1}{c}{Ba$_4$In$_2$Br$_2$O$_6$}   & \multicolumn{1}{c}{ICSD}            & 81878       \\ \mr
  \multicolumn{1}{c}{Na$_4$Cd$_4$Sb$_4$S$_{12}$}  & \multicolumn{1}{c}{ICSD}            & 183395      \\ \mr
  \multicolumn{1}{c}{Ba$_2$Sn$_2$F$_8$}      & \multicolumn{1}{c}{ICSD}            & 166207      \\
  \end{tabular}
\end{table}

\begin{figure}[tb]
\centering
\includegraphics[width=\linewidth]{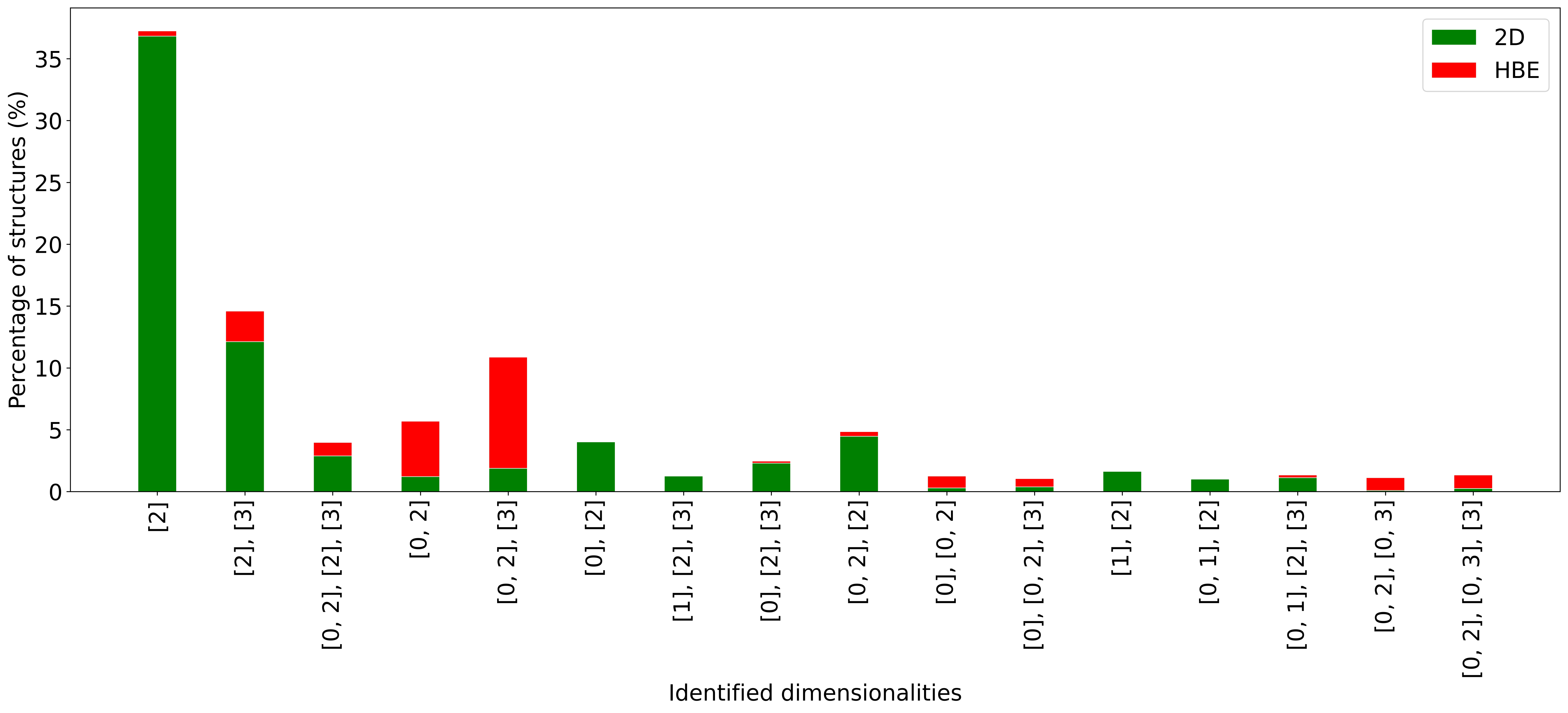}
\caption{Evaluation of the performance of geometrical screening step (\emph{LowDimFinder}) with respect to accurate results on the binding energies calculated using DFT. 
This figure is very similar to Fig. 2 in the main text, in particular we refer to its caption for the meaning of the bar labels. However, here we do not show the performance of our model, but only of the \emph{LowDimFinder} geometrical step. Structures are color-coded based on their binding energy as later computed by DFT: exfoliable structures (labeled as 2D) are shown in green, high-binding energy structures (labeled as HBE) are shown in red.}
\label{fig:SI-Barplot-detail}
\end{figure}

\clearpage
\bibliographystyle{iopart-num}
\bibliography{bibliography}


\end{document}